\documentclass[11pt,a4paper]{article} 

  \usepackage[
  hyperindex,
  pagebackref,
  colorlinks,
  linkcolor=blue,
  menucolor=blue,
  pagecolor=blue,
  urlcolor=blue,
  citecolor=blue,
  pdfcreator={LaTeX}
  ]{hyperref}

  \usepackage{thumbpdf}
  \usepackage{url}

  \usepackage[dvips]{graphicx}
  \DeclareGraphicsExtensions{.eps,.jpg,.mps,.png}

\usepackage{fancyhdr} 

\begin{document} 


\pagestyle{myheadings} 
\markboth{Starlab preprint KN-2005-11-12}{STARLAB Preprint   KN-2005-11-12} 
\begin{titlepage} 
\vspace{1.in} 

\begin{center} 
\ 

\ 

{\Huge\bf Complex networks in brain electrical activity \\[0.4in]   } 

\begin{figure}[h] 
\begin{center} 
\includegraphics[width=5cm]{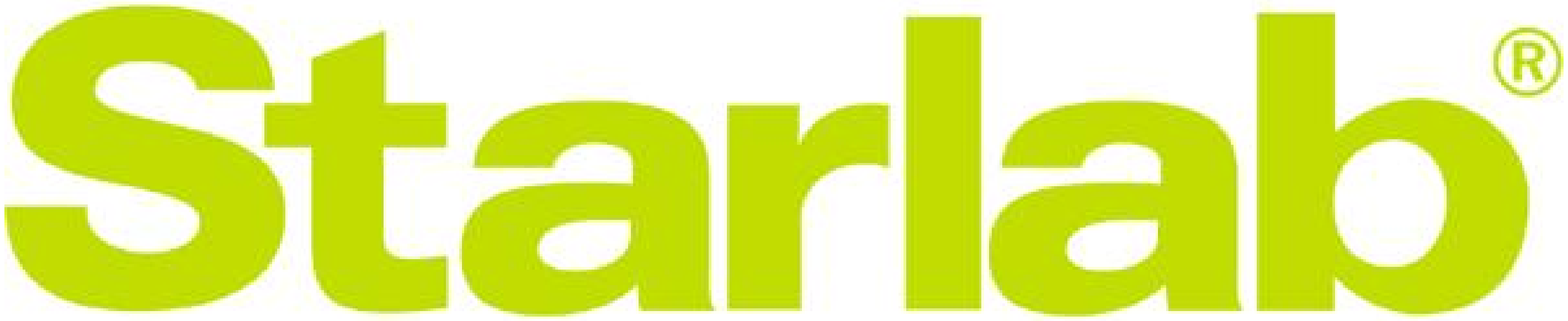} 
\end{center} 
\end{figure}

{\large\bf Starlab Knowledge Nugget KN-2005-11-15, Kolmogorov   project} \\ 
{\large \bf Status: Starlab Preprint}\\ 
{\large \bf Project Initiated: April 7th 2005}\\ 


\vspace{0.5cm} \noindent 
{\normalsize {\bf Authors:} G. Ruffini\footnote{Contact:   giulio.ruffini@starlab.es},  C. Ray, J. Marco,  L. Fuentemilla, C.   Grau} \\ 
\noindent {\normalsize {\bf Date:} Nov 15th 2005}\\ 
\noindent {\normalsize {\bf Version:} 1.8}\\ 

        {\small\it Starlab Barcelona, S.L.}\\ 
        {\small\it Edifici de l'Observatori Fabra, C. de   l'Observatori s.n.}\\ 
{ \small\it Muntanya del Tibidabo, 08035 Barcelona, Spain \\  http:// starlab.es } 

\end{center} 

\vspace{.1in} 
\begin{center} 
{\large\bf Summary} 
\end{center}


{\small 
This paper is the result of work carried out  at Starlab  in   collaboration with the Neurodynamics Laboratory of the U. of   Barcelona (UBNL) focusing on complex networks analysis of EEG data   (provided by UBNL). The approach is inspired by the work in \cite  {Eguiluz} with fMRI data and represents a follow up on earlier   efforts on analysis of ERP/MMN data using tomography and  independent  component analysis to characterize brain connectivity  and spatial  funcionalization \cite{Marco}.   Multichannel EEG  measurements are first processed to obtain 3D voxel activations  using the tomographic algorithm LORETA. Then, the correlation of  the current intensity activation between voxel pairs is computed to  produce a voxel cross-correlation coefficient matrix. Using several  correlation thresholds, the cross-correlation matrix is then  transformed into a network connectivity matrix and analyzed. The  resulting analysis highlights significant differences between the  spatial activations associated with Standard and Deviant tones,  with interesting physiological implications. When compared to  random data networks, physiological networks are more connected,  with longer links and shorter path lengths. Furthermore, as  compared to the Deviant stimulus case, Standard data networks are  consistently more connected, with longer links and shorter path  lengths---consistent with a ``small worlds'' character. The  comparison between both networks shows that areas known to be  activated in the MMN wave are connected. In particular, the  analysis supports the idea that supra-temporal and inferior frontal  data work together in the processing of the differences between  sounds by highlighting an increased connectivity in the response to  a novel sound. 
} 

\end{titlepage} 
\ 

\clearpage

\clearpage 

\begin{center} 
{\huge \bf  Complex networks in brain electrical activity} 
\end{center} 
{\em 
\begin{center} 
{\large  G. Ruffini$^1$, C. Ray$^{1,2}$, J. Marco$^{1,3}$,  L.   Fuentemilla$^3$, C. Grau$^3$  }\\ 
$^1$Starlab, C. de l'Observatori, s/n, 08035 Barcelona, Spain,  http:// starlab.es \\ 
$^2$Physics Department, St. Mary's College, Moraga, CA, USA \\ 
$^3$Neurodynamics Laboratory, Department of Psychiatry and  Clinical  Psychobiology, University of Barcelona, Spain 
\end{center} 
} 

\begin{abstract} 

{\small 

A major challenge for neuroscience is to  map and analyze  the   spatiotemporal patterns of activity of the large neuronal  populations  which are believed to be responsible for information  processing in  the human brain. In this paper,  we analyze the  complex networks  associated with brain electrical activity in a  specific experimental  context.  Our approach uses multichannel EEG  data analysis to  characterize the  spatial connectivity of the  brain. Multichannel EEG  measurements are first processed to obtain  3D voxel activations   using the tomographic algorithm LORETA.  Then, the correlation of the  current intensity activation between  voxel pairs is computed to  produce a voxel cross-correlation  coefficient matrix. Using several  correlation thresholds, the  cross-correlation matrix is then  transformed into a network  connectivity matrix and analyzed. 
To study a specific example, we selected data from an earlier   experiment focusing on the Mismatch Negativity (MMN) brain wave--- an  Event-Related Potential (ERP)---because it gives an   electrophysiological index of a ``primitive intelligence''  associated  with auditory pattern  and change detection in a  regular auditory  pattern. EEGs have an exceptional millisecond  temporal resolution,  but appear to result from mixed neuronal  contributions whose spatial  location and relationships are not  fully understood. 
Although that experimental setup was not optimal for the purpose  of  tomography and network analysis (only 30 electrodes were  used),  the  resulting  analysis has already detected significant  differences  between the spatial activations associated with  Standard and Deviant  tones, with interesting physiological  implications. As a cross-check,  we have also analyzed the networks  from randomly generated data. When  compared to random data  networks, physiological networks are more  connected, with longer  links and shorter path lengths. Furthermore,  as compared to the  Deviant stimulus case, Standard data networks are  consistently  more connected, with longer links and shorter path  lengths--- consistent with a ``small worlds'' character.  The  clustering  index is rather high in general, especially in comparison  to an  equivalent random network. The departure from randomness is   extreme at high thresholds.  On the other hand, the clustering  index  does not appear to discriminate clearly the Standard and  Deviant  networks, while relative low correlation thresholds appear  to be more  discriminating. Yet, when compared to random or Deviant  data,  Standard data networks appear to fragment more easily, with  the  number of sub-networks increasing at larger thresholds. 
On the other hand, the comparison between both networks shows that   areas known to be activated in the MMN wave are connected. In   particular, the analysis   supports the idea that supra-temporal  and  inferior frontal data work together in the processing of the   differences between sounds by highlighting an increased  connectivity  in the response to a novel sound. } 

\end{abstract} 
\clearpage 
\ 

\clearpage 

\section{Introduction} 
The availability of large amounts of data and  increasingly  powerful  computers is enabling the analysis of complex systems  with many   highly interconnected elements. Examples of such  systems abound both   natural and artificial, and include the cell,  the brain, society, the  internet, the behavior of crowds and the  economy  to name a few \cite {Albert, Valverde}.  The brain is  perhaps the most interesting and  complex of all such systems, with  hundreds of billions of highly  interconnected neurons. Complex  networks theory provides a tool for  the analysis of the  connectivity of such complex systems, where  interesting phenomena  are emergent from a high number of highly  interacting elements. It  is therefore rather natural that this  methodology will be put to  practice intensely to study information  processing in the brain at  different scales. \medskip 

While a complete analysis of the brain at cellular level  is  beyond  present capabilities, we can analyze the function and  connectivity of  the brain at relatively small scales. One of the  early views in  neuroscience is that specific areas of the brain  are specialized in  performing specific tasks. This working  hypothesis has led to the  discovery of several functional  properties of distinct areas by  lesional, intracortical  and  neuroimaging  studies.  However,  functional localization cannot be  uncovered by the analysis of  anatomical localization alone, as  such an approach would not  uncover  functional interaction  integration of different brain areas.  Functional integration  refers to an effective connectivity, that is,  it depends on both  connectivity dynamism and upon a model of  interactions \cite {Friston2005}. A more general model postulates the  processing of  information through cooperation of different localized  regions.   \medskip 

Although several powerful functional approaches to image-specific   neuronal assemblies while the human brain is functioning are   available today \cite{Toga}, they all have a limited  application   niche.  Some are invasive, while others are based on an indirect   index of brain computing---such as metabolic or hemodynamic   measurements which are blind to millisecond phenomena (Positron   Emission Tomography, PET; and functional Magnetic Resonance, fMRI,   for instance). In contrast to these methods, electromagnetic scalp   brain activity techniques (Electroencephalography, EEG,  Event-  Related Potentials, ERP, and their magnetic counterparts) have the   advantages of being non-invasive, of being closely related to the   traffic of ions through the dendritic ion channels that are  believed  to support cerebral computing \cite{Kandel}, and of  having an  exceptional temporal resolution (in the order of a  millisecond).  While these signals seem to be very rich in  information, information  extraction has been traditionally elusive  \cite{Nunez}. In  particular, the coarse spatial resolution of EEG  has limited its  impact. Traditionally, studies on this topic have   focused in the  connectivity of scalp electrodes. However, recent  advances in the  search for the sources of the brain electrical  activity (the so- called inverse problem) have provided a foothold  for the use of EEG  to study the spatial  connection between  distant areas of the brain. 
\medskip 

In \cite{Eguiluz} fMRI human brain data was analyzed using a  complex  networks approach, and evidence was found for a scale free  behavior  of the derived functional networks---basically the degree  probability  distribution  obeyed a power law for high degrees.   This paper seeks  to analyze, in a similar vein, EEG data. EEG data  is very  complementary to fMRI data, which is metabolic in nature.  EEG is a  non-invasive technique with a high temporal resolution  and a medium  spatial resolution, as we discuss below. While fMRI  measurements are  readily available as 3D fields, EEG data is only  measured at the  scalp.  Since our interest is  focused on 3D  localization of the  networks associated with large neuronal  groups,  the first step in  our approach  is to  tomographically  ``decode'' electrode data and  map it into 3D voxel activation  data---the derived networks are then  to reflect 3D spatial  connectivity patterns. \medskip

In this paper we describe a ``complex networks''  methodology for   electrophysiology  and apply it to a relatively simple set of data   obtained in an Event Related Potential protocol with 13 subjects  and  collected with 30 EEG channels. While is certainly true that  30  electrodes are barely sufficient to obtain good tomographic  results,  we aim here to carry out a first simple test of the  approach. The  technique can be applied with more sophisticated  data sets, and the  envisioned applications are many. 
\medskip 

Our approach will be applied to a set of  single trial ERP data in  a  Mismatch Negativity (MMN) paradigm \cite{Natanen78}. MMN is a  wave  that appears 100-200 ms after the incoming of a tone that  breaks the  regularity of an stream of regular tones.  In order to  probe the  structure of a particular functional brain network, we  will consider  in this study the impact produced by an  environmental change. The  brain can be studied as a pattern  modeling tool, where inputs from  the environment  transduced by   body sensors are analyzed. The  capability to  model ambient  inputs  is crucial for the survival of  higher organisms. Neural  networks responsible for this task   determine whether incoming  information needs to deeply alter  brain  dynamics (robustness) or  to adjust (responsiveness), even  dramatically, in order to  effectively respond \cite{Bar-Yam2004}.  \medskip 

MMN reflects the capacity of the system to detect a sudden   environmental change\cite{Natanen92} which constitutes a  fundamental  ability to ensure the survival of the organism \cite  {Tiitinen,Jaaskelainen}. MMN constitutes an auditory ERP that has   been suggested to be a measure of auditory information pre- attentive  processing \cite{Paavilainen,  Javitt, Naatanen1997},  as  one of the  first stages of gating information to consciousness  \cite {Jaaskelainen}. The MMN system has been suggested to reflect  a  ``primitive intelligence'' \cite{Naatanen2001} that provides an  in  vivo simplified model for studying abstract brain processing  and  related memory mechanisms. 
The MMN wiring network includes several brain areas, as concluded    from studies  using diverse techniques (including EEG, fMRI and  PET).  It has been described to relate mostly to both supra- temporal  hemispheres \cite{Alain2001,Rosburg2004, Jaaskelainen,  Kircher2004,  Muller2002} and frontal areas \cite{Giard1990,  Opitz2002, Muller2002,  Doeller2003,  Liasis2001}, with a  predominance of the right  hemisphere, and also parietal sources  \cite{Kasai1999, Levanen1996}.  A step forward would be to describe  interrelations between them  including possibly their sequentially  participation in the time  domain \cite{Marco}. However, a complex  networks analysis can lead to  a better  understanding of the  connections (links) between the  components (nodes)---that is, the  functional connectivity.


\section{Experimental setup} 
\begin{figure}[b!] 
\vspace{-5cm} \ 

{\centering \hspace{3.cm} 
\includegraphics[width=4.in] {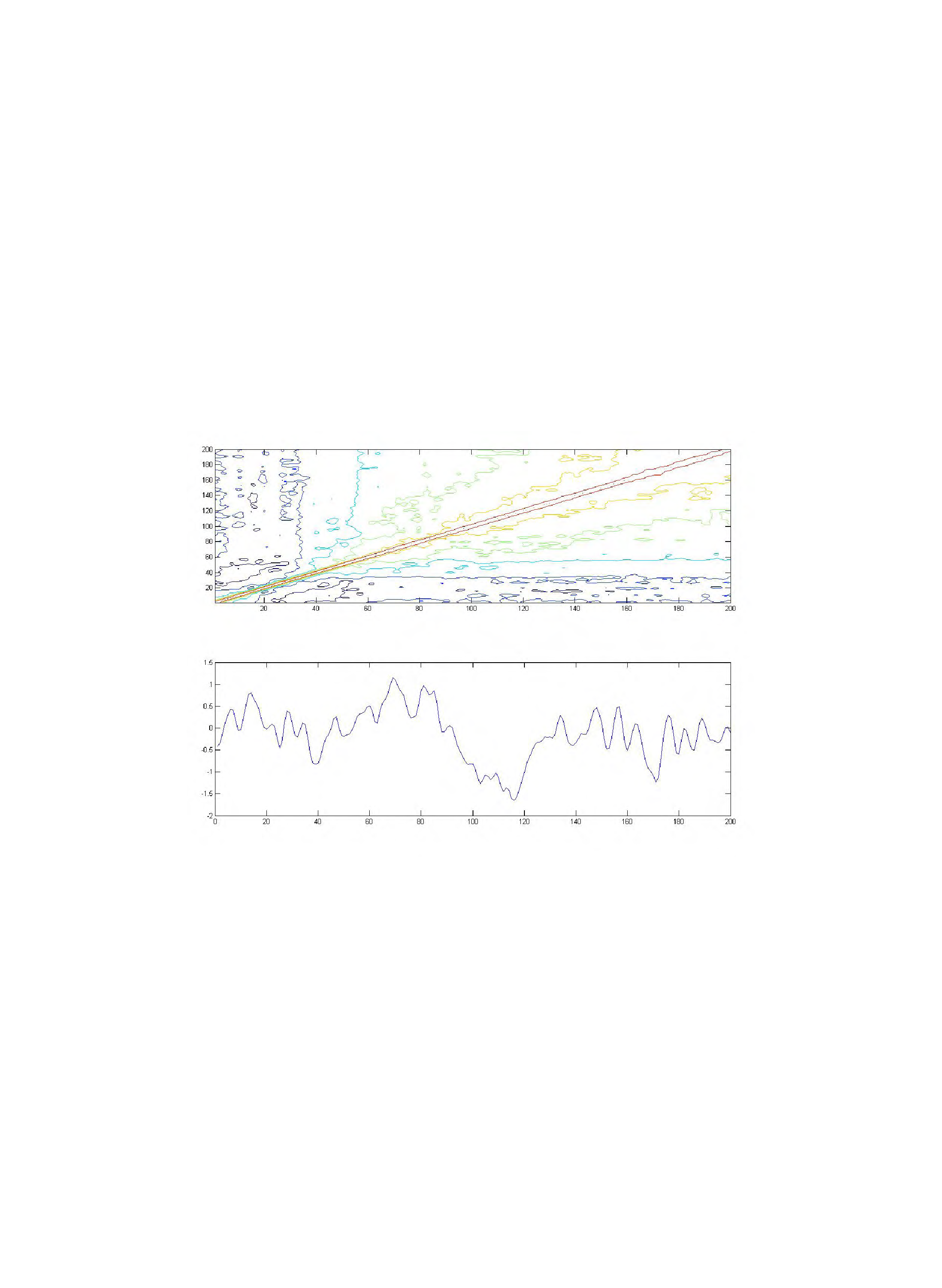} } \vspace{4.2cm} 

\caption{Top: Autocorrelation function of the ERP for different  times  (x and y axis denote time steps of 2 ms). This is computed  from 300  trials (one subject), for electrode CZ. Note that the  process is not  stationary, as the statistics depend on time  reference. Coherence  times change are short before stimulus onset  (at the 50 time step  mark) and increase monotonically. Bottom:  Average voltage over 300  trials.  \label{lineup} } 
\end{figure}

In this work we have applied our analysis to a set of previously   obtained and well studied data.  The  paradigm described in \cite  {Grau,Marco}  was designed  to study brain electrical response  from  the incoming of a novel stimulus in a background of know  stimuli.  Stimuli (85 dB SPL) consisted of pure sine-wave tones of  700 Hz, with  a duration of 75 ms (Standard tone) or 25 ms (Deviant  tone), with 5  ms of fall/rise time. Trains of 3 tones were  presented to subjects  binaurally. The first tone of trains was  Standard (p=0.5) or Deviant  (p=0.5), while the other two tones  were Standard. The inter-stimulus  time was 300 ms, and the  temporal separation between trains was 400  ms. A total of 400  stimuli trains were presented randomly. 
As can be seen in Figure~\ref{lineup}, the signal (here for a  single  electrode) becomes self-coherent after stimulus onset.  Figure~\ref{lineup} displays a  plot of the inter-trial correlation  coefficient function of the voltage time series for one electrode  \cite{Ramsay1997}, 
\begin{equation} 
A_v(t_1, t_2)= {\hbox{Cov}(t_1, t_2) \over \sigma_v(t_1)\,  \sigma_v (t_2)}. 
\end{equation} 
The coherence time of the signal, $t_c$, at a point $t$ is the  time  it takes for the $A_v(t+t_c, t-t_c)$ to fall to $1/2$. The  coherence  time can be visualized by inspecting $A_v$ along lines  of the form $ (t_1,t_2)=(t,t)+t_c (-1,1)=(t-t_c, t+t_c)$. \medskip 

Sixteen healthy subjects (mean age 39$\pm$11 years) participated  in  the study after having given their written consent. The  subjects were  instructed to ignore auditory stimuli while they   performed an  irrelevant visual task (watching TV with the sound  off). EEGs  (bandpass 0-100 Hz) were recorded with a SynAmps  amplifier (Neuroscan  Inc) at a sampling rate of 500 Hz. A total of  30 electrodes were  used: eighteen followed the 10-20 system  without O1 and O2 (FP1, OZ,  FP2, F7, F3, FZ, F4, F8, T3, C3, CZ,  C4, T4, T5, P3, PZ, P4, T6), and  twelve more electrodes (FC1, FC2,  FT3,FT4, M1, M2, IM1, IM2, TP3,  TP4, CP1 and CP2), all of them  referenced to a nose electrode. Two  extra electrodes were used to  record vertical and horizontal ocular  movements. Recordings were  notch-filtered at 50 Hz. Epochs exceeding $ \pm$100µV in EEG or in  electro-oculogram were automatically rejected.  Bandpass filtering 
(0.1-30 Hz) was performed, obtaining epochs of 400 ms, 100 ms  before  stimulus to 300 ms after it. Three subjects were excluded  from the  rest of the study because they did not have an  identifiable MMN wave.  In what follows, electrode voltage time  series will be identified by $ \phi_q(t_i)$, where the $q$ index  refers to the electrode number, and  $t_i$ to the time of sampling  events.

\section{Computation of the spatial Connectivity Matrix from EEG data} 
The network is constructed from the  ERP data in three steps which  we  now describe. 
\begin{enumerate} 
\item  Low Resolution Tomography (LORETA) \cite{Pasqual-Marqui} is   employed to estimate the current densities within the brain from  the  measured electrode voltages on the scalp.  According to our  general  understanding of the origin of bioelectric signals \cite {plonsey},  the potentials we observe on the body surface are due  to ionic  current sources at the neuron membranes.  As a result of  these  currents a charge density exists. More specifically, charges  arise  where impressed current density field lines are born or die.  These  charges are the source of the electrical potential at the  scalp we  measure. Therefore, given an accurate model of  this  forward process,  we expect to be able to invert the operation and  determine the cause  (currents) from the effect (voltages). 
There is an important difficulty, however, as there is not a  unique  inverse solution: many patterns of current density can lead  to the  same measured electrode voltage pattern. 
For this reason it is not possible to exactly determine the  current  density from the electrode voltages. 
The LORETA algorithm picks out a current density solution with an   adaptive resolution by searching for the smoothest of all possible   current density solution maps $J_n$ that satisfy boundary  conditions  derived from the position of cortical grey matter and  the hippocampus  of the Talairach human brain model. 
In the past five years, this tomographic approach has been used in   several neuroscience studies (see for example \cite{Kounios,  Mulert,  Pizzagalli, Gomez, Marco}). 
The LORETA version used in this study searched for the sources of   activation in 2394 voxels distributed in the Talairach human brain   \cite{Pasqual-Marqui}.  \medskip 

It will be useful to write here, in a simplified form, the   tomographic equations. For each time step, the forward map $K_{qn} $,  mapping current densities into electrode voltages, can be  written as 
\begin{equation} 
\phi_q=\sum_n K_{qn} J_n, 
\end{equation} 
and the inverse transformation $T_{nq}$, which is the inverse  chosen  by the LORETA algorithm, can be written as 
\begin{equation} 
\tilde J_n=\sum_q T_{nq}\phi_q. 
\end{equation} 
We note here that we use a simplified notation, since $J_n$ is   actually a 3-vector. The LORETA inverse map gives an estimate of  the  current densities $\tilde J$ from the electrode voltages. 
Note that $KT$ is by definition of inverse the identity, while $TK $  is in general  not  the identity. 
The ``activation'' of a voxel is defined to be the magnitude of  the  estimated current density, 
\begin{equation} 
I_n=|| \tilde J_n||=|| T_{nq}\phi_q||. 
\end{equation} 
For each subject, for each time step of each 400 (400 ms long)  trials  and for each subject,  the electrode data is transformed  into a voxel  current density magnitude using LORETA, producing  2394 voxel current  density magnitude time series $I_n(t_i)$ for  each trial. 
There are 100 Standard and 100 Deviant trials for each subject. 

\item 
The second step is the computation of the inter-voxel correlation   coefficient. 
The correlation coefficient  ${c_{nm}}$ between voxels $n$ and $m$   for a single trial, is defined to be the correlation coefficient  of  the activations over time. 
\begin{equation} 
c_{nm}= {\langle I_nI_m \rangle_t -  \langle I_n \rangle_t\langle  I_m  \rangle_t\over \sigma_n\sigma_m} = \langle S_nS_m \rangle_t 
\end{equation} 
where $\langle \cdots \rangle_t$ denotes a time average, $\sigma_n=  \sqrt{\langle I_n^2 \rangle_t-\langle I_n \rangle_t^2}$ is the   standard deviation of the activation, and $S_n(t_i)={[I_n(t_i) -   \langle I_n\rangle_t ]/ \sigma_n}$ is the demeaned and normalized   activation signal. 
This coefficient can be interpreted as the cosine of the angle   between the two activation time series vectors. 
The activation correlation coefficient used here is defined to be  the  average over all 100 trials of $c_{nm}$.  We will use the  notation $  \bar c_{nm}$ for the average activation correlation  coefficient matrix.

\item 
As in \cite{Eguiluz}, the links of the network are determined from   the activation correlation coefficient matrix  by applying a   correlation threshold: when the absolute value of the correlation   between voxels $n$ and $m$ is greater than the threshold $r_c$,  the  voxels are linked, otherwise the voxels are not linked. 
This defines the connectivity matrix, with entries equal to 1  (voxels  are linked) or 0 (not linked), 
\begin{equation} 
A_{nm} = \cases{1&if $\vert \bar c_{nm}\vert>r_c$ and $n\ne m$ \cr  0&  otherwise} 
\end{equation} 
\end{enumerate} 
Therefore, the nodes of the brain activation network are  identified  here with the voxels, while the links are specified by  the  connectivity matrix after thresholding (as in \cite{Eguiluz}).  In the  data set that is analyzed here a network is constructed for  each  subject for both deviant and standard conditions and for  each  threshold.  This gives a total of 168 networks.


\section{Analysis of correlations induced by the tomographic  inverse  transformation} 
One of the problems we encountered here is that the tomographic   processing induces artificial correlations in the voxel  activations.  This is easily understood by a counting argument. In  our test data  set there were 30 data sources (the electrodes) and  almost three  thousand voxels, and the two spaces are related by a  linear  transformation. It is clear, then, that there will be  correlations in  the voxel activations even if the electrode data  is randomly  generated and fully de-correlated. Today it is  possible to work with  up to 256 electrodes, and this will  certainly increase the effective  tomographic resolution.  Nevertheless, even with 256 electrodes the  problem will remain and  needs to be carefully analyzed. Here we  provide only a preliminary  analysis; we will leave a more detailed  study of this issue to a  future publication.  In this paper we  acknowledge that the network  structure obtained is affected by the  resolution problem, and for  this reason we will focus only on the  differences between two  experimental setups. To first order,  inversion induced correlation  and network biases will then cancel. 
\medskip 

While the true correlation coefficient matrix is not accessible,  we  can estimate the inversion induced correlations by analyzing  two  scenarios. The first one, already discussed, asks the question  of  what correlations are induced by the inversion if the electrode  data  is fully decorrelated. This scenario is  appropriate for the  analysis  of decorrelated noise impact.  Table~\ref {table:randelect} provides a  summary of the network parameters  from a Montercarlo simulation. 
\begin{table}[t!] 
{\centering 
\begin{tabular}{|c||c|c|c|c|c|c|c|c|}  \hline 
$r_c$     & $\langle k \rangle$  &  $\langle C\rangle$ &  $\langle  C_ {rand}\rangle$  &   $\langle d \rangle$& $ \langle l\rangle $ &   $  \langle l_{rand}\rangle $   &$l_{max}$  &  $T$ \\  \hline \hline 
0.4    &286    &0.67&0.12     &37    &2.7    &1.8 &6     & 1 \\ \hline 
0.5    &194    &0.69&0.08    &31    &3.4    &1.9 &7     & 1\\  \hline 
0.6    &137    &0.69&0.06    &26    &4.2    &2.0 &9     & 1\\ \hline 
0.7    &97    &0.70&0.04    &22    &5.3    &2.1 &12     & 1\\ \hline 
0.8    &64    &0.69&0.03    &19    &7.0    &2.2 &16     & 1\\ \hline 
0.9    &33    &0.67&0.01    &15    &11.3    &2.5 &28     & 5 \\\hline 
\end{tabular} 
\caption{Network parameters (see Section~5) for random uncorrelated  electrode data.  All number are dimensionless except for$\langle d  \rangle$, which is  in mm.  The values for clustering index and  average path length for  an equivalent random network are also  given.\label{table:randelect}} 
} 
\end{table}

\medskip 

The second is the scenario in which the original voxel signals are   decorrelated: after forward mapping and then inversion,  the   resulting voxel cross-correlations  can be used as an inversion   induced background correlation threshold which can be used to   estimate the inversion impact on the obtained connectivity  matrix.   This inversion induced correlation coefficient can in  principle be  computed from the tomographic equations. However, we  recall here that  the experimental connectivities have been  obtained from the current  activations (current density norm) in  order to simplify data  analysis. Nevertheless, for future  reference we will compute here the  inversion current correlation  coefficient  although is not directly  applicable to the present  analysis. \medskip 

Suppose that the true current sources are given by $J_n$. It can   easily be shown that if the true current sources are fully   decorrelated, the cross-correlation of the estimated currents is   simply proportional to $R^t R$. After forward ($K$) and back ($T$)   mapping, the estimated currents are given by 
\begin{equation} 
\tilde J_n = \sum_m (TK)_{nm} J_m \equiv \sum_m  R_{nm} J_m . 
\end{equation} 
We begin by calculating the average value of the estimated currents, 
\begin{eqnarray} 
\langle \tilde J_i \rangle_t &=& \langle \sum_n R_{in} J_n \rangle_t  \nonumber \\ 
&=&  \sum_n R_{in} \langle J_n\rangle_t \nonumber \\ 
&=&  (RA)_i , 
\end{eqnarray} 
where the vector $A$ is defined by $A_{n} \equiv \langle  J_n   \rangle_t$. The cross-correlations are given by 
\begin{eqnarray} 
\langle \tilde J_i \tilde J_j \rangle_t &=& \langle \sum_n R_{in}  J_n \ \ \sum_m R_{jm} J_m \rangle_t \nonumber  \\ 
&=& \sum_{n,m} R_{in}R_{jm}\langle  J_n J_m \rangle_t \nonumber \\ 
&=& \sum_{n,m} R_{in}\langle  J_n J_m \rangle_t R_{mj}^t \nonumber \\ 
&=&(RB R^t)_{ij} , 
\end{eqnarray} 
where the matrix $B$ is defined by $B_{nm} \equiv \langle  J_n J_m   \rangle_t$, that is, the real cross-correlations. 
Similarly, the standard deviations are 
\begin{eqnarray} 
\sigma_i &=& \sqrt{\langle \tilde J_i^2 \rangle_t - \langle \tilde   J_i \rangle_t^2} \nonumber \\ 
&=& \sqrt{(RB R^t)_{ii} - (R A)_i^2}. 
\end{eqnarray} 
The cross correlation is given by 
\begin{eqnarray} 
\tilde C_{ij} &=& {\langle  \tilde J_i \tilde J_j \rangle_t -   \langle  \tilde J_i \rangle_t\langle\tilde J_j \rangle_t \over   \sigma_i \sigma_j } \nonumber  \\ 
&=& {(RB R^t)_{ij} - (RA)_i(RA)_j \over \sqrt{(RB R^t)_{ii} - (R A)  _i^2}\sqrt{(RB R^t)_{jj} - (R A)_j^2} } . 
\end{eqnarray} 
Therefore, we see that the obtained correlation coefficient is a   complex function of the biophysical correlations and the inversion   matrices. As an example, consider the situation where the $J_n$  are  orthonormal and with zero mean, so that  $\langle   J_n J_m  \rangle_t= \delta_{nm}$ and $\langle   J_n  \rangle_t=0$.  Then,   $B=1$ and $A=0 $ and 
\begin{eqnarray} 
\langle \tilde J_i \rangle_t &=& (RA)_i=0  ,\\ 
\langle \tilde J_i \tilde J_j \rangle_t &=& (RR^t)_{ij},   \\ 
\sigma_i &=& \sqrt{(RR^t)_{ii} }, 
\end{eqnarray} 
and 
\begin{equation} 
\tilde C_{ij} = {(R R^t)_{ij}  \over \sqrt{(RR^t)_{ii} (RR^t)_  {jj} }  } \equiv \tilde  C_{ij}^0 . 
\end{equation} 
This quantity is in some sense the zero-point correlation in the   estimated currents, since it is the correlation in the estimated   currents when the actual currents having zero correlation.  For  this  reason we expect this to be an indicative measure of the  correlations  induced by the  inversion transform. \medskip


\begin{figure}[t] 
\begin{center} 
\includegraphics[width=5 in]{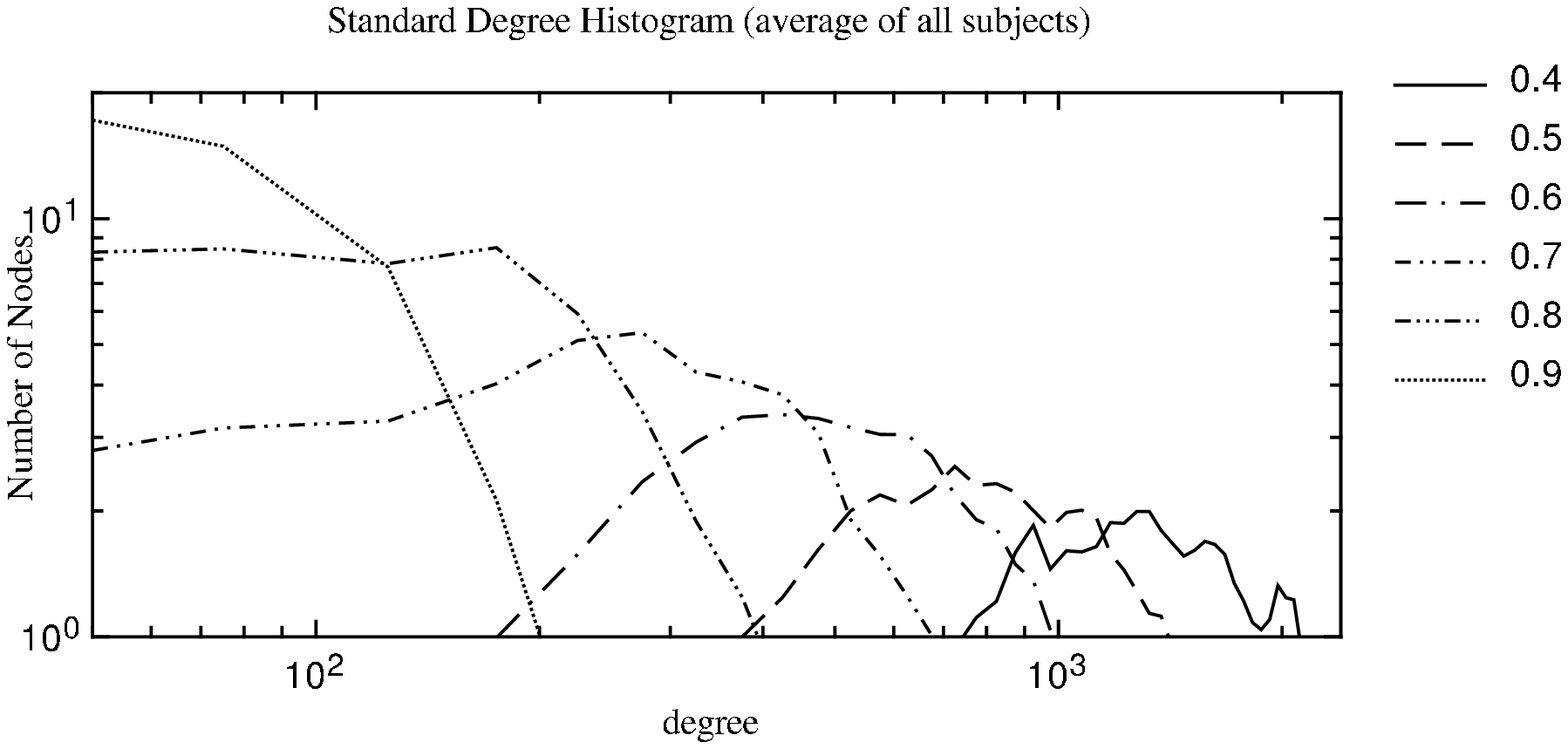} 

\includegraphics[width=5 in]{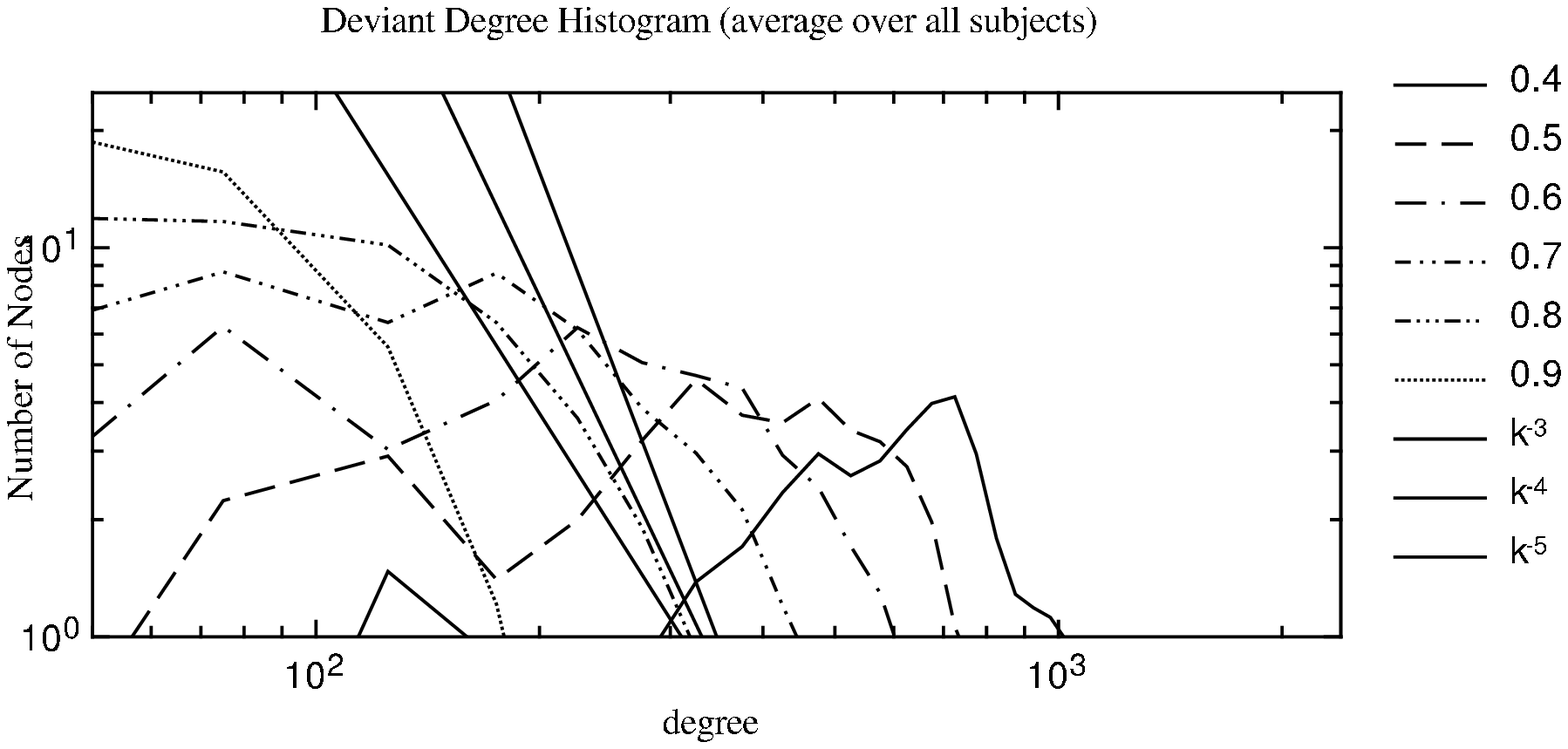} 

\end{center}\vspace{-0.7cm} 

\caption{Degree distributions (node count as a function of degree)   for the two experimental conditions (averaged over all subjects),  and  for different thresholds. \label{histograms}} 
\end{figure} 
\section{Network analysis} 
Once a network is defined by the connectivity matrix, we can study   measures associated with the network\cite{Albert,Eguiluz}. The   following measures are computed for each node of a network: 
\begin{itemize} 
\item $k_n$ is the degree of node $n$, the number of nodes linked  to  node $n$: 
\begin{equation} 
k_n = \sum_i A_{ni}=\sum_i A_{ni}A_{ni}=\sum_i A_{ni}A_{in}= A_{nn}^2 
\end{equation} 
\item $C_n$ is the cluster coefficient for node $n$, the ratio of  the  number of links between the neighbors of node $n$ and the  maximum  possible number of links between the neighbors. 
With $\nu_n=\{m \vert A_{nm}=1\}$ the set of neighbors of node $n$  we  can write the cluster coefficient as follows, 
\begin{equation} 
C_n = {\sum_{i,j\in\nu_n} A_{ij} \over k_n(k_n-1)}= {A_{nn}^3  \over  k_n(k_n-1)}. 
\end{equation} 
Note that the second form is equivalent because the restricted sum  of  $A_{ij}$ is the same as the unrestricted sum of $A_{ni}A_{ij}A_ {jn}$  since either $A_{ni}$ or $A_{jn}$ is zero for the added  elements of  the sum and both are 1 for the original elements of  the sum. 
The cluster coefficient is not defined for nodes with less than  two  neighbors. 
\item $L_{nm}$ is the path length between nodes $n$ and $m$, the   minimum number of links required to travel through the network  from  node $n$ to node $m$. 
The path length is undefined if no path between the nodes exists. 
\end{itemize} 
Based on these measures, the following parameters have been  computed  for each subject's connectivity matrix $A$ and threshold  $r_c$, and  for each of the experimental conditions (Standard or  Deviant tone): 
\begin{itemize} 
  \item $N$, is the number of nodes with at least one link. This   measure  is not shown, since it does not deviate much (less than 1\ 
\item $\langle k \rangle$ is the average node degree for a  particular  network.  The notation $\langle \cdots \rangle$  indicates a network  average, 
\begin{equation} 
\langle k \rangle = {1\over N}\sum_n k_n = {1\over N} {\rm Tr}[A^2]. 
\end{equation} 
\item $\langle L \rangle$ is the average path length, an indicator  of  the connectedness of the network. 
Pairs of nodes without a connecting path are not included in the   average. 
\item $L_{max}$ is the maximal path length (the network  perimeter).  For a random network it is approximated  by $L_{max}=  \ln N / \ln  \langle k \rangle$ \cite{Fronczak2004}. 
  \item $\langle C \rangle $ is the average cluster coefficient,  an  indicator of the fraction of completed sub-networks. 
As described in \cite{Albert}, a common property of social  networks  is that cliques form, representing circle of friends or  acquaintances  in which every member knows every other member. 
This  tendency to cluster is quantified by the clustering coefficient. 
For an equivalent random network $\langle C \rangle = \langle k   \rangle /( N-1)$. 
  \item $T$ is the number of {\em tribes}: we define a tribe as   disjoint sub-network, i.e., one disconnected from the rest of the   network. This can be easily computed from $L_{mn}$, by creating an   associated matrix with ones everywhere except for disconnected  voxel  pairs (where the value is zero), and computing its rank. 
  \item $\langle d \rangle$ is the average link physical length  (in  mm), which is not a topological measure  but relevant  nonetheless in  the present study (although as pointed out in \cite {Eguiluz} cortex  folding makes its interpretation more difficult), 
\end{itemize} 
For the computation of difference statistics given in Table \ref  {table:stan-dev}, the difference of these measures for the  Standard  and Deviant networks is computed for each subject. Then  the average  and standard deviation are computed.   \medskip 

While more sophisticated tools in complex networks analysis are   available, searching for the frequency of motifs, we concentrate  in  this first effort on a ``classical'' complex networks  description.  \medskip 

\begin{figure}[t] 
\begin{center} 
\hspace{0.3cm}\includegraphics[width=14cm]{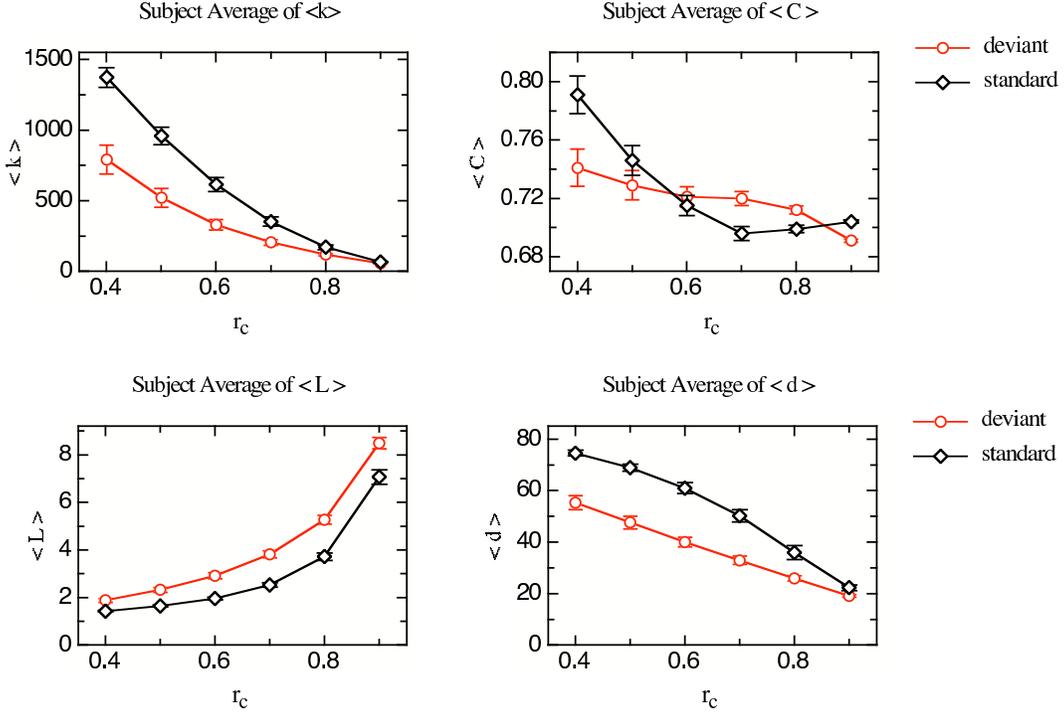} 
\end{center} \vspace{-0.7cm} 

\caption{Average network parameters as a function of correlation   threshold for all subjects and for the two experimental  conditions--- Standard and Deviant. The error bars reflect the   uncertainty of the  average values. \label{avepara}} 
\end{figure} 

\begin{figure}[t] 
\begin{center} 
\includegraphics[width=5 in]{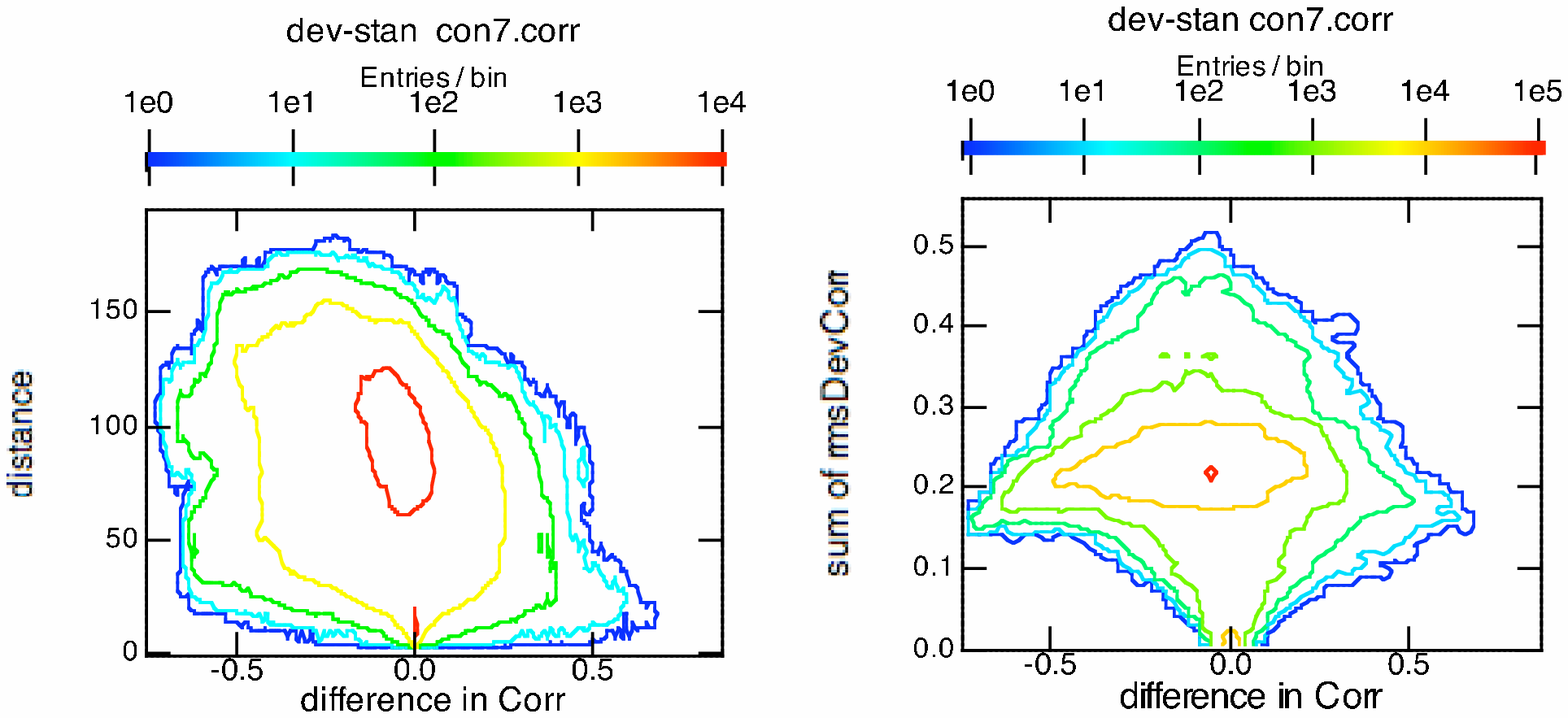} 
\end{center} \vspace{-0.7cm} 

\caption{Left: Link count as a function of Correlation coefficient   difference (Deviant minus Standard) and link length  (mm)  for one   subject (the results are similar for all subjects). Right: as a   function of standard deviation of the correlation coefficient over   100 trials. \label{link-distribution}} 
\end{figure} 

Results for the  characteristics of the networks associated with  the  brain electrical response to a Standard and Deviant tone  without  tomographic corrections are plotted in Figure~\ref {avepara}. The  degree distributions are provided in Figure~\ref {histograms}, and the  correlation coefficient difference  distribution (link count) is  provided in Figure~\ref{link- distribution}. As can be seen there are  marked differences between  conditions in almost all studied  parameters. The evaluation of the  statistical differences between  both conditions is summarized in  Table~\ref{table:stan-dev}, while  the individual Standard and  Deviant parameters are given in the  summary Table~\ref {table:both}. The number of links ($ \langle $k$  \rangle$),  average  physical distance ($\langle$d$\rangle$), and path  length  (averaged and maximum) present significant differences in the   networks associated with the Standard and  Deviant cases for all  the  studied correlation coefficients. The cluster coefficient and  number  of tribes are nearly constant over the range of thresholds  studied.   The number of tribes does increase for the largest  threshold in the Standard case, although in all cases the network  was formed by one large tribe with more than 99\% of the nodes and  the remaining nodes were  not connected to each other. As can  be  seen in Table~\ref{table:ttest}, the differences are  statistically  significant.

\begin{table}[b!] 
{\centering 
\begin{tabular}{|c||c|c|c|c|c|c|c|c|}  \hline 
$r_c$   & $\langle k \rangle$  &  $\langle C\rangle$ &  $\langle C_  {rand}\rangle$&   $\langle 
d \rangle$& $ \langle l\rangle $&   $ \langle l_{rand}\rangle $ &  $l_ {max}$  &  $T$ \\  \hline 
\hline 
    0.4 &   1395 (255)& 0.80 (0.04)&  0.58 & 75 (4)& 1.4 (0.1)&  1.5   &2.8 (0.4)&  1 (0)\\  \hline 
    0.5&    986 (222)&  0.75 (0.03)&0.41  &  70 (5)& 1.6 (0.1)&  1.5  &3.6 (0.5)&  1 (0)\\  \hline 
    0.6&    637 (170)&  0.72 (0.02)&   0.27& 62 (6)& 1.9 (0.2)&    1.6&4.6 (0.5)&  1 (0)\\  \hline 
    0.7&    365 (120)&  0.70 (0.02)&   0.15& 51 (9)& 2.5 (0.3)&    1.7&6.2 (0.8)&  2 (0)\\  \hline 
    0.8&    175 (64)&   0.70 (0.02)&  0.07 & 37 (10)&3.6 (0.5)&  1.9  &9.5 (1.3)&  2 (0)\\  \hline 
    0.9&    68 (13) &   0.70 (0.00)&   0.03& 23 (4)& 6.9 (1.0)&   2.2  &21.7 (3.2)& 9 (1)\\  \hline 
\end{tabular} 
\vspace{0.2cm} 

\begin{tabular}{|c||c|c|c|c|c|c|c|c|}  \hline 
$r_c$   & $\langle k \rangle$  &  $\langle C\rangle$ &  $\langle C_  {rand}\rangle$&   $\langle 
d \rangle$& $ \langle l\rangle $&   $ \langle l_{rand}\rangle $ &  $l_ {max}$  &  $T$ \\  \hline 
\hline 
    0.4&   \hspace{0.2cm} 723 (303)&  0.73 (0.04)&0.30 &   54 (8)&   1.9 (0.3)& 1.6& 3.8 (0.5)&  1 (0)\\  \hline 
    0.5&    477(198)&   0.73 (0.02)&  0.20&  46 (8)& 2.4 (0.3)&  1.7&  5.0 (0.6)&  1 (0)\\  \hline 
    0.6&    306 (111)&  0.72 (0.02)&   0.13& 39 (6)& 3.0 (0.4)&  1.7&  6.5 (0.8)&  1 (0)\\  \hline 
    0.7&    192 (60)&   0.72 (0.01)&   0.08& 32 (5)& 3.9 (0.5)&  1.9&  8.5 (0.7)&  1 (0)\\  \hline 
    0.8&    113 (24)&   0.71 (0.01)&   0.05& 25 (3)& 5.4 (0.6)&  2.0&  12.5 (1.0)& 1 (0)\\  \hline 
    0.9&    54 (9)&         0.69 (0.01)&   0.02& 19 (2)& 8.6 (0.8)  &2.3&  22.5 (2.6)& 2 (1)\\  \hline 
\end{tabular} 
\caption{Top: Network parameters for Standard data. Bottom:  Network  parameters for Deviant data. All numbers are dimensionless  except for$ \langle d \rangle$, which is in mm. The standard  deviation of the  data is shown in parenthesis---note that the mean  uncertainty is  smaller by the square root number of subjects. The  values for  clustering index and average path length for an  equivalent random  network are also given. \label{table:both}} 
} 
\end{table}

\begin{table}[t!] 
{\centering 
\begin{tabular}{|c||c|c|c|c|c|c|}  \hline 
$r_c$     & $\langle k \rangle$  &  $\langle C\rangle$ &   $\langle  d  \rangle$& $ \langle l\rangle $ & $l_{max}$  &  $T$ \\  \hline   \hline 
    0.4&    671 (276)&    0.07 (0.05)&    21 (8)&    -0.5 (0.2)&     -1.0 (0.7)&    0 (0)\\  \hline 
    0.5&    559(189)&    0.03 (0.03)&    24 (7)&    -0.7 (0.3)&     -1.4 (0.6)&    0 (0)\\   \hline 
    0.6&    331 (122)&    0.00 (0.03)&    23 (7)&    -1.1 (0.4)&     -1.8 (0.8)&    0 (0)\\  \hline 
    0.7&    173 (85)&    -0.02 (0.02)&    19 (7)&    -1.4 (0.4)&     -2.3 (0.7)&    1 (0)\\  \hline 
    0.8&    63 (48)&    -0.01 (0.01)&    11 (8)&    -1.7 (0.5)&     -2.9 (1.3)&    1 (0)\\   \hline 
    0.9&    13 (11) &        0.01 (0.01)&    4 (4)&    -1.7 (0.9) &    -0.8 (3.7)&    7  (2)\\  \hline 
\end{tabular} 
\caption{Average difference of Standard and Deviant network  measures  (i.e., the subject by subject measured difference and  then  statistics). All numbers are dimensionless except for$\langle  d  \rangle$, which is in mm. Standard deviation of the data is  shown in  parenthesis---note that the mean uncertainty is reduced  by one over  the square root of the number of subjects. \label {table:stan-dev} } 
} 
\end{table} 

\begin{table} 
{\centering 
\begin{tabular}{|c||c|c|c|c|c|c|}  \hline 
$r_c$   & $\langle k \rangle$  &  $\langle C\rangle$ &   $\langle 
d \rangle$& $ \langle l\rangle $ & $l_{max}$  &  $T$ \\  \hline 
\hline 
    0.4&    5.87$^{***}$& 3.9$^{***}$& 7.98$^{***}$ & -5.99$^{***} $&  -5.38$^{***}$& 0\\  \hline 
    0.5&    5.92$^{***}$& 2.6$^{*}$&8.96$^{***}$&-6.96$^{***} $&-6.50 $^{***}$&0\\  \hline 
    0.6&    5.62$^{***}$&0.17&9.12$^{***}$&-7.68$^{***}$&-6.57$^ {***} $&1\\  \hline 
    0.7&    4.47$^{***}$&-3.14$^{**}$&6.78$^{***}$&-8.31$^{***}  $&-7.31$^{***}$&5.20$^{***}$\\  \hline 
    0.8&    2.94$^{**}$&-3.69$^{**}$&3.55$^{**}$&-7.71$^{***} $&-6.21 $^{***}$&11.08$^{***}$\\  \hline 
    0.9&    2.87$^{**}$&4.54$^{***}$&2.86$^{**}$&-4.44$^{***}  $&-0.71&15.94$^{***}$\\  \hline 
\end{tabular} 
\caption{Student's t-test statistics for the comparison between   Standard and Deviant data. Here  * p$<$0.05, ** p$<$0.01, *** p$<  $0.001. \label{table:ttest}} 
} 
\end{table}


\medskip 

\begin{figure}[h] 
\begin{center} 
\includegraphics[width=3.5 in]{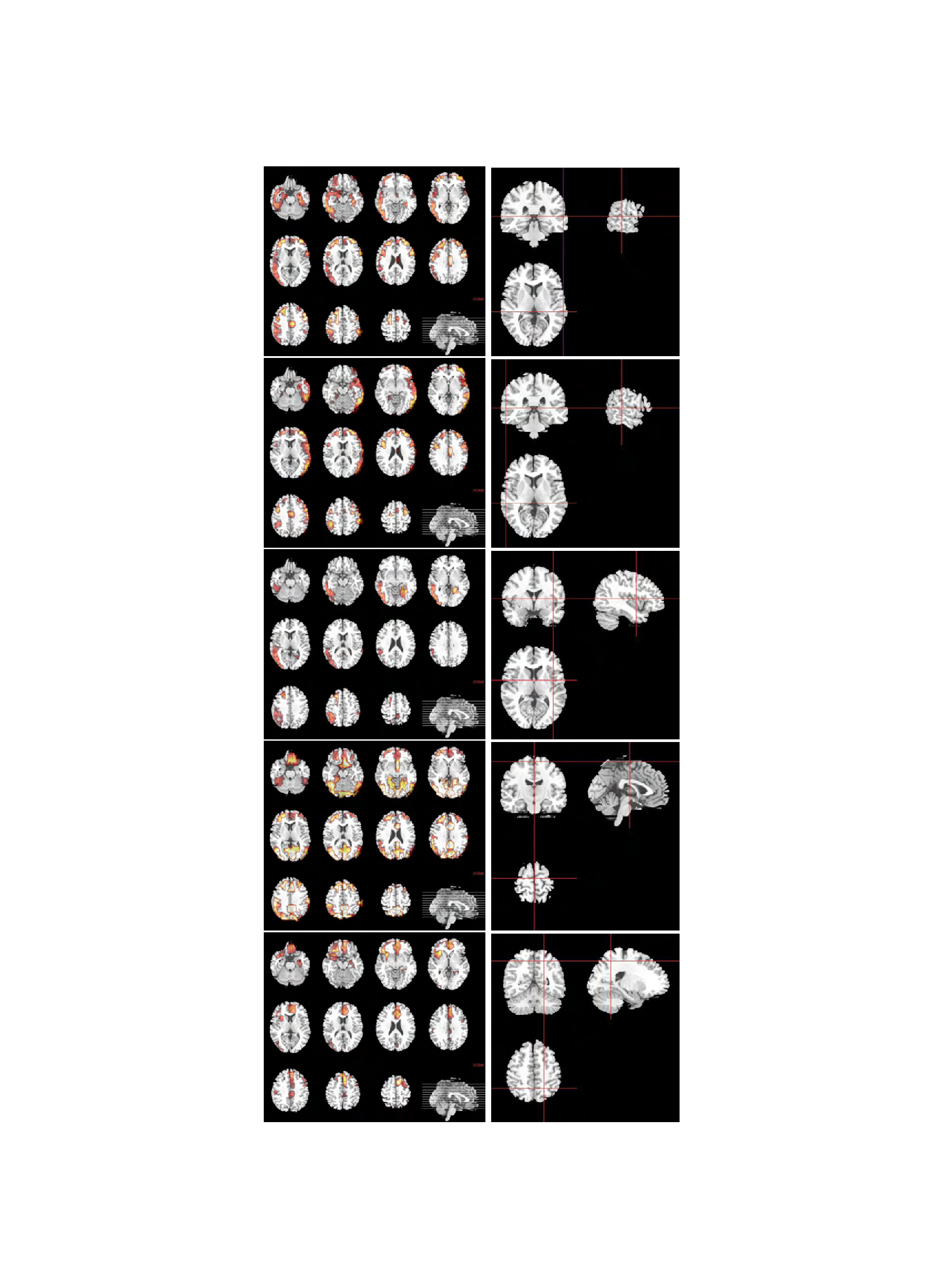} 
\end{center}\vspace{-0.5cm} 

\caption{Deviant minus Standard correlations for the right   supratemporal voxel (top), left supratemporal voxel, right  inferior  frontal voxel, central frontal voxel and right inferior  parietal  voxel (bottom).\label{fig:con} } 
\end{figure}

Results for uncorrected data on voxels activated in the Mismatch   Negativity process are shown in the Figure~\ref{fig:con}, which  shows  the correlations of all voxels with respect to selected  reference  voxels. Reference voxels are selected based on the  results obtained  in \cite{Marco} with the same data set. As it can  be seen, several  important differences exist between networks for  the Standard and  Deviant responses. We can state the following: 
\begin{itemize} 
\item {\em  Relative to the Deviant, the Standard network is more   connected (with significantly more links at any given threshold),   with longer links, with a shorter average and maximal path length. } 
\item {\em The increase in number of links in the Standard  condition  is greater at lower correlations, while at high  thresholds the  Standard network fragments more easily  relative to  the Deviant  network.  } 
\end{itemize} 

This seems to imply that the  Standard-Deviant difference is   characterized by  a relative increase of long scale links at lower   correlations in the Standard network, and also to a relative  increase  of short links at higher correlations in the Deviant  network, which  avoid fragmentation. Thus, we can say that the  Standard network has a  more ``small world'' character \cite {Watts1998} than the deviant--- especially at lower correlation  coefficient thresholds. Both have  shorter path lengths and  perimeters than the random electrode data  network (see Table~\ref {table:randelect}). These can also be compared  to Random network  theory \cite{Albert,Fronczak2004}. The cluster  coefficient for a   random graph (Erdos-Renyi) is simply given by  $ \langle C \rangle  = \langle k \rangle /( N-1)$, while according to  \cite {Fronczak2004}, the average path length is given by 
\begin{equation} 
\langle L \rangle={\ln N- \gamma \over \ln (\langle k \rangle)} . 
\end{equation} 
Although to some extent the same is true for the Deviant network,  the  results for the Standard case indicate, especially for low   thresholds, a similar path length as would be expected from a  random  network, while the clustering index appears to be a factor  of 2  greater. These results are similar to those obtained in \cite  {Salvador2005} and rather different than those in \cite{Eguiluz}.   They are also similar to analyses of the cat cortex \cite  {Scannell1999} and the macaque visual cortex \cite{Hilgetag2000}.    The obtained  networks with low thresholds display ``small world''   properties, with high clustering and short path lengths  \cite  {Watts1998}. This is especially true of the Standard network. It  is  interesting to note here that the clustering index is not very   sensitive to the threshold, while the same is not true of the path   length. This is consistent with the long but decreasing link  lengths  with increasing thresholds. Relatively low correlation  links provide  long scale connectivity. 
\medskip 

With regards to scale free behavior, it is not readily apparent in   either one, and in this sense our results are similar to those in   \cite{Hilgetag2000, Salvador2005, Scannell1999} and different from   those in \cite{Eguiluz}. The perturbation due to the inversion-  induced correlations might  mask such behavior to some extent,   however, so it is not possible to make a definitive statement. It   appears as if at higher correlations there Standard/Deviant   differences are smaller---and thus of less physiological  interest--- with the exception of the number of tribes  (disconnected sub- networks). As mentioned, there is a clear  difference in the  histograms of the networks especially at lower  correlations (see  Figure~\ref{histograms}). \medskip 

The Standard network is in essence much more connected and  clustered  at long scales, while at short scales the Deviant  network appears to  be more robust.

\section{Discussion} 
In this paper we have provided a preliminary study of  the   statistical characteristics of brain networks associated with the   brain electrical activity.  We have addressed two fundamental   questions. The first one is if complex network analysis can be  used  to discriminate, in a statistical significant way, between  the chosen  experimental conditions, and the stability of the  measures across  subjects. We have discussed the problems  associated with tomographic  inversion in preparation for network  analysis, which made the outcome  of the study less certain.  The  second question is if the results  be  interpreted  physiologically.   What can we learn about the  spatial   organization of brain information processing using a complex  networks  approach? 

\medskip 

With regards to the first question, the results of this study show   that  network analysis can be used to discriminate between two   different brain functional states with inter-subject stability,   showing a potential relevance in the analysis EEG/MEG and ERPs/ ERFs.  Another interesting result comes from the fact that,  although  parameters are highly dependent on the threshold used,  the  differences between Standard and Deviant data remain for the  average  number of links, average and maximal path length and  average link  physical length. 
\medskip 

The first result that arises from the present study is that the   number of links per voxel (node) in the Standard condition is   significantly greater than in the Deviant condition---especially  at  lower correlation coefficient thresholds and longer scales.  This  result is, in principle, unexpected due to the fact that the  brain  electrical response to a Deviant tone elicits a greater  activity  mainly in supratemporal and frontal cortices \cite {Natanen92,  Opitz2002} indexed by MMN. However this result is  consistent in all  studied subjects and for all thresholds, and  appears to be robust.  This phenomenon appears to be related to   the average link physical  length, greater  also in the Standard  than in Deviant conditions,  indicating a relative increase in  longer scale connections in the  Standard condition. The  combination of both parameters (number of  links and average  connection distance) suggests a first physiological  result: 
\begin{itemize} 
\item {\em The Standard condition presents  higher connectivity at   longer spatial scales and a strong ``small world'' character.} 
\end{itemize} 

This is consistent with the classical view of the brain's  electrical  response to Deviant data. Supratemporal areas that are  processing the  characteristics of sound in the Standard conditions  have to recruit  nearby areas (also supratemporal ones) to perform  the extra activity  needed by the automatic detection of  difference. Our results could  also be interpreted in the framework  of \cite{Jaaskelainen}, where it  is postulated that MMN is not a  mechanism itself, but a non-reduced  N1. Following this theory on  MMN, supratemporal areas involved in the  generation of the N1  would be the same as the areas that generate  MMN. Links in the  Deviant condition would be shorter because the main  areas  implicated in the production of MMN would be closer than in the   attenuated N1. However, although Deviant data present less  numerous  and shorter links than Standard data, the Deviant network  connects in  a more efficient way all the cerebral areas as  reflected by the  greater number of ``tribes'' in the Standard  network at higher  thresholds. \medskip 

As mentioned, these results are similar to those obtained in \cite  {Salvador2005} and rather different than those in \cite{Eguiluz}---  both working with fMRI data at much longer time scales. 
\medskip 

In a previous analysis of the same data set \cite{Marco} using   Independent Component Analysis tomography, we found that the MMN  is  mediated by orchestrated activity of several spatially  distributed  brain sources located in the temporal, frontal, and  parietal  areas,which activate at distinct time intervals. When  studying some  of these areas  several aspects became apparent. It  was found that a  right and a left supratemporal structure and the  right inferior  frontal cortex seem to be connected in the MMN  condition (Deviant  minus Standard) indicating that these  structures could be related  with the same brain function.  Traditionally, the participation of  temporal structures in the MMN  ERP was associated with the analysis  of the sound, whereas frontal  structures were associated with the so- called ``call to  attention'' \cite{Natanen1990, Opitz2002,Doeller2003}. However, in  recent years, it has been  found that these structures could be  implied in the same functioning,  as reflected by the fact that  they could be present in the same  independent component \cite {Marco}. Moreover, primate studies have  revealed that right  inferior frontal structures are activated in  response to a sound.  The present study, which indicates that these  areas are more  connected in the response to a novel sound compared to  Standard  data, is consistent with previous findings and supports the  idea  that supratemporal and inferior frontal data work together in  the  processing of the differences between sounds.  It is also clearly  observable in Figure~\ref{fig:con} that right frontal and parietal  areas are linked together in deviant detection, which agrees with  the hypothesis that both areas are connected in attentional  processes \cite{Kasai1999, Levanen1996}.  In this framework, it is  stated that a spatially distributed set of neuronal groups that are  activated coherently and are part of the same representation form  an assembly \cite{Engel2001}. In other terms, it could be described  as distributed local networks of neurons transiently linked by  reciprocal dynamic connections, which supports functional  integration \cite{Varela2001}. 
\medskip

The application of this approach could range from the basic  studies  determining the properties of networks associated with  event related  potentials or electroencephalography, to the study  of pathological  brain electrical responses, to biometrics.  However, more studies have  to be made in order to compare the  information provided by an  electrophysiology  complex networks  approach with information  provided by other functional techniques  (such as fMRI) and  theoretical information to clearly validate  this method. In  particular, we plan to analyze elsewhere in more  detail the impact of  inversion in the network structure, as well  as study other variants  for the construction of the networks.

\section*{Acknowledgments} 
This work has been partly funded by the Starlab Kolmogorov program   under the auspices of the FURNET network. G. Ruffini would like to   thank Ed Rietman for valuable discussions, and in particular for   pointing out the relevance of \cite{Eguiluz}. The authors wish to   acknowledge support received from the Generalitat for support in  the  NECOM group (SGR2005)  and the EU FP6 Sensation Integrated  Project  (FP6-507231)  for partial support. Finally, the authors  thank Carles  Escera and  Maria Dolores Polo (U. Barcelona) for  data collection.

\end{document}